\RequirePackage{fix-cm}
\documentclass[twocolumn,epjc3,longtitle]{svjour3}       % onecolumn (second format)
\smartqed  % flush right qed marks, e.g. at end of proof
\journalname{Eur. Phys. J. C}
\usepackage{amsfonts}
\usepackage{amssymb}
\usepackage{multirow}
\usepackage{soul}
\usepackage{graphicx}
\def\Journal#1#2#3#4{#1 \textbf{#2}, (#3) #4} 
\def\NIMA{Nucl.\,Instrum.\,Methods A}
\def\NIMB{Nucl.\,Instrum.\,Methods B}
\def\NPB{Nucl.\,Phys. B}

\def\APL{Applied\,Phys.\,Lett.}
\def\PLB{Phys.\,Lett.  B}

\def\PRL{Phys.\,Rev.\,Lett.}
\def\PRA{Phys.\,Rev. A}
\def\PRD{Phys.\,Rev. D}
\def\PRC{Phys.\,Rev. C}
\def\EPJ{Eur. Phys. J.}

\def\NDS{Nuclear Data Sheets}
\def\JLT{J.\,Low\,Temp.\,Phys.}
\def\JINC{J.\,Inorg.\,Nucl.\,Chem.}

\def\Re{$^{187}$Re}
\def\Ho{$^{163}$Ho}

\def\agre{AgReO$_4$}
\def\sinx{SiN$_x$}
\def\Tr{$^3$H}
\def\mn{$m_\nu$}
\def\mnsq{$m_\nu^2$}
\def\mug{$\mu$g}
\def\mum{$\mu$m}

\def\mus{$\mu$s}

\def\taur{$\tau_{R}$}
\def\tm{$t_M$}
\def\de{$\Delta E$}
\def\fwhm{$_{\mathrm{FWHM}}$}
\def\Aec{$A_\mathrm{EC}$}

\def\fpp{$f_{pp}$}
\def\Nev{$N_{ev}$}
\def\ero{Er$_2$O$_3$}
\def\hoo{Ho$_2$O$_3$}
\def\Ho{$^{163}$Ho}

\def\Er{$^{162}$Er}

\def\unimib{Dipartimento di Fisica, Universit\`a di Milano-Bicocca, Milano, Italy}
\def\infnmib{Istituto Nazionale di Fisica Nucleare (INFN), Sezione di Milano-Bicocca, Milano, Italy}
\def\unige{Dipartimento di Fisica, Universit\`a di Genova, Genova, Italy}
\def\infnge{Istituto Nazionale di Fisica Nucleare (INFN), Sezione di Genova, Genova, Italy}
\def\lngs{Laboratori Nazionali del Gran Sasso (LNGS), INFN, Assergi (AQ), Italy}

\def\infnroma{Istituto Nazionale di Fisica Nucleare (INFN), Sezione di Roma 1, Roma, Italy}
\def\nist{National Institute of Standards and Technology (NIST), Boulder, Colorado, USA}
\def\ILL{Institut Laue-Langevin (ILL), Grenoble, France}
\def\PSI{Paul Scherrer Institut (PSI), Villigen, Switzerland}
\def\lisboa{Multidisciplinary Centre for Astrophysics (CENTRA-IST), University of Lisbon, Lisbon, Portugal}
\def\caltech{Jet Propulsion Laboratory, California Institute of Technology, Pasadena, California, USA }

\begin{document}

% \title{The HOLMES experiment}
\title{HOLMES}
\subtitle{The Electron Capture Decay of \Ho\ to Measure the Electron Neutrino Mass with sub-eV sensitivity}
% \classification{23.40.Bw; 14.60.Pq; 07.20.Mc; 29.40.Vj}
% <Replace this text with PACS numbers; choose from this list:                 \texttt{http://www.aip..org/pacs/index.html}>}

\author{B.\,Alpert\thanksref{nist}, %{address={\nist}}
M.\,Balata\thanksref{lngs}, %{address={\lngs}, 
D.\,Bennett\thanksref{nist}, %{address={\nist}, 
M.\,Biasotti\thanksref{unige,infnge}, %{address={\unige},  altaddress={\infnge}, 
C.\,Boragno\thanksref{unige,infnge}, %{address={\unige},  altaddress={\infnge}, 
C.\,Brofferio\thanksref{unimib,infnmib}, %{address={\unimib},  altaddress={\infnmib}, 
V.\,Ceriale\thanksref{unige,infnge}, %{address={\unige},  altaddress={\infnge}, 
D.\,Corsini\thanksref{unige,infnge},
M.\,De Gerone\thanksref{unige,infnge}, %{address={\unige},  altaddress={\infnge}, 
P.\,K.\,Day\thanksref{caltech},
R.\,Dressler\thanksref{psi}, %{address={\PSI}, 
M.\,Faverzani\thanksref{unimib,infnmib}, %{address={\unimib},  altaddress={\infnmib}, 
E.\,Ferri\thanksref{unimib,infnmib}, %{address={\unimib},  altaddress={\infnmib}, 
J.\,Fowler\thanksref{nist}, %{address={\nist}, 
F.\,Gatti\thanksref{unige,infnge}, %{address={\unige},  altaddress={\infnge}, 
A.\,Giachero\thanksref{unimib,infnmib}, %{address={\unimib},  altaddress={\infnmib}, 
J.\,Hays-Wehle\thanksref{nist},
S.\,Heinitz\thanksref{psi}, %{address={\PSI}, 
G.\,Hilton\thanksref{nist}, %{address={\nist}, 
U.\,K\"oster\thanksref{ill}, %{address={\ILL}, 
M.\,Lusignoli\thanksref{infnroma}, %{address={\uniroma}, 
M.\,Maino\thanksref{unimib,infnmib}, %{address={\unimib},  altaddress={\infnmib}, 
J.\,Mates\thanksref{nist}, %{address={\nist}, 
S.\,Nisi\thanksref{lngs}, %{address={\lngs}, 
R.\,Nizzolo\thanksref{unimib,infnmib}, %{address={\unimib},  altaddress={\infnmib}, 
A.\,Nucciotti\thanksref{e1,unimib,infnmib}, %{address={\unimib},  altaddress={\infnmib}, 
G.\,Pessina\thanksref{infnmib}, %{address={\infnmib}, 
G.\,Pizzigoni\thanksref{unige,infnge}, %{address={\unige},  altaddress={\infnge}, 
A.\,Puiu\thanksref{unimib,infnmib}, %{address={\unimib},  altaddress={\infnmib}, 
S.\,Ragazzi\thanksref{unimib,infnmib}, %{address={\unimib},  altaddress={\infnmib}, 
C.\,Reintsema\thanksref{nist}, %{address={\nist}, 
M.\,Ribeiro Gomes\thanksref{lisboa}, %{address={\lisboa}, 
D.\,Schmidt\thanksref{nist}, %{address={\nist}, 
D.\,Schumann\thanksref{psi}, %{address={\PSI}, 
M.\,Sisti\thanksref{unimib,infnmib}, %{address={\unimib},  altaddress={\infnmib}, 
D.\,Swetz\thanksref{nist}, %{address={\nist}, 
F.\,Terranova\thanksref{unimib,infnmib}, %{address={\unimib},  altaddress={\infnmib}, 
J.\,Ullom\thanksref{nist}}%{address={\nist}}
\thankstext{e1}{e-mail: angelo.nucciotti@mib.infn.it}
\institute{\nist\label{nist}
\and \lngs\label{lngs}
\and \unige\label{unige}
\and \infnge\label{infnge}
\and \unimib\label{unimib}
\and \infnmib\label{infnmib}
\and \caltech\label{caltech}
\and \PSI\label{psi}
\and \infnroma\label{infnroma}
\and \ILL\label{ill}
\and \lisboa\label{lisboa}
}

\date{Received: date / Revised version: date}

\maketitle

\abstract{
The European Research Council has recently funded HOLMES, a new experiment to
directly measure the neutrino mass. HOLMES will perform a calorimetric
measurement of the energy released in the decay of \Ho. The calorimetric
measurement eliminates systematic uncertainties arising from the use of external
beta sources, as in experiments with beta spectrometers. This measurement was
proposed in 1982 by A. De Rujula and M. Lusignoli, but only recently the
detector technological progress allowed to design a sensitive experiment. HOLMES
will deploy a large array of low temperature microcalorimeters with implanted
\Ho\ nuclei. The resulting mass sensitivity will be as low as 0.4\,eV. HOLMES
will be an important step forward in the direct neutrino mass measurement with a
calorimetric approach as an alternative to spectrometry. It will also establish
the potential of this approach to extend the sensitivity down to 0.1\,eV. We
outline here the project with its technical challenges and perspectives. 
}

%%%%%%%%%%%%%%%%%%%%%%%%%%%%%%%%%%%%%%%%%%%%
%% MAINMATTER
%%%%%%%%%%%%%%%%%%%%%%%%%%%%%%%%%%%%%%%%%%%%

\section{Introduction}
Assessing the neutrino mass scale is one of the major challenges in 
today's particle physics and astrophysics. Although neutrino oscillation 
experiments have clearly shown that there are at least three neutrinos 
with different masses, the absolute values of these masses remain unknown. 
Neutrino flavor oscillations are sensitive to the difference between the 
squares of neutrino mass eigenvalues and have been measured by solar, 
atmospheric, reactor, and accelerator experiments \cite{oscillations}. 
Combining such results, however, does not lead to an absolute value for 
the eigenmasses, and a dichotomy between two possible scenarios, 
dubbed "normal" and "inverted" hierarchies, exists. 
The scenario could be complicated further by the hypothetical existence of 
additional "sterile" neutrino eigenvalues at different mass scales \cite{SterileNu}. 

The value of the neutrino mass has many implications, from cosmology 
to the Standard Model of particle physics. In cosmology the neutrino mass affects the 
formation of large-scale structure in the universe. In particular, 
 neutrinos tend to damp structure clustering, before they have cooled sufficiently 
to become non-relativistic, 
with an effect that is dependent on their mass \cite{nu-structureformation}. 
In the framework of $\Lambda$-CDM cosmology, the scale dependence of 
clustering observed in the Universe can, indeed, be used to set an 
upper limit on the neutrino mass sum in the range between about 0.2 and 
1\,eV \cite{nu-CMB},  
although this value is strongly model-dependent. In 
particle physics, 
a determination of the absolute scale of neutrino masses would 
help shedding light on the mechanisms of mass generation. 

For years, laboratory experiments based on the study of proper 
nuclear processes have been used to directly measure the neutrino 
masses. In particular, single beta decay has been, historically, 
the traditional and most direct method to investigate the electron 
(anti)neutrino mass \cite{betaendpoint}. Neutrinoless double beta decay has also been 
suggested as a powerful tool to measure the electron neutrino mass, 
although the decay itself, and thus its efficacy at measuring the 
neutrino mass, is indeed dependent on the assumption that 
the neutrino is a Majorana particle \cite{bb-Majorana}.

To date, the study of the beta decay of $^3$H using electrostatic 
spectrometers has been the most sensitive approach, yielding an 
upper limit on the electron anti-neutrino mass of 2.2\,eV \cite{MainzTroitsk}. 
In the near future, the new experiment KATRIN will analyze the $^3$H 
beta decay end-point with a much more sensitive electrostatic 
spectrometer and an expected statistical sensitivity of about 
0.2\,eV \cite{KATRIN}.
However KATRIN reaches the maximum size
and complexity practically achievable for an experiment of its type and no further
improved project can be presently envisaged.

Project8 \cite{Project8} is a new science program with the ambition to reach a \mn\ sensitivity beyond the one of KATRIN by performing Cyclotron Radiation Emission Spectroscopy of \Tr\ decay in a  \emph{source\,=\,detector} configuration which avoids the systematic uncertainties arising from an external electron source. Despite being a quite young technique
it is very promising and 
first results are impressive \cite{Project8-2014}.

An alternative to spectrometry is provided by calorimetric measurements in
which the beta source is embedded in the detector. In this configuration the energy
emitted in the decay is entirely measured by the detector, except for the fraction
taken away by the neutrino. 
This approach eliminates both the problematics of an external source and the systematic uncertainties coming from the  excited final states.
One perfect way to make a calorimetric
measurement is to use low temperature thermal detectors \cite{cryodet}, where all excitations
from an interaction are allowed to decay to near thermal equilibrium thus producing
a temperature rise of the detector that can be measured by a sensitive thermometer
to determine the deposited energy. The measurement is then free from
the systematic effects induced by possible energy loss in the source and it is not affected
by uncertainties related to decays to excited final states. If no meta-stable
states exist in the material all the decay energy is integrated and collected in a
short time regardless of its origin: X-ray, Auger electron or recoil. Since calorimeters
detect all the occurring decays, the source activity must be limited to avoid
spectral distortions and background at the end-point due to pulse pile-up.

Since the fraction of decays in a given energy interval \de\ below the end-point $Q$ scales only as $(\Delta E/Q)^3$, the limitation on the
statistics may be partially balanced by using as beta source \Re, the beta-active nuclide with the lowest known transition energy ($Q\approx2.5$\,keV).
Because of \Re\ natural isotopic abundance of about 63\%, a metallic rhenium crystal can be used as detector absorber for calorimetry. Rhenium metal undergoes a superconducting transition at $T_c = 1.7$\,K, so that the electronic specific heat vanishes when operated at very low $T/T_c$. This allows the use of a relatively large amount of metallic rhenium (about 1\,mg). However metallic rhenium exhibits two major difficulties: the low specific activity of about 1\,Bq/mg and the relative slowness of the thermalization process. Also dielectric compounds can be used, but they do not completely circumvent the metal problems, since they have even lower specific activities and they face an incomplete thermalization of the deposited energy, which limits the achievable energy resolution at 2.5\,keV to about 20\,eV. 
Several investigations have been carried out on microcalorimeters for a calorimetric neutrino mass measurement with \Re.
About 10 years ago, two small scale pilot experiments were carried out with thermal detectors containing \Re: the MANU \cite{manu} and MIBETA \cite{mibeta} experiments. Both experiments collected a statistics of about $10^7$ decays, yielding a limit on \mn\ of about 15 eV at 90\% CL. The sensitivity of these experiments was limited by statistics and detector performance, while systematic effects were relatively small. Since then, the project ``Microcalorimeter Arrays for a rhenium Experiment (MARE)'' has been facing the demanding task of improving and scaling up those pioneer experiments using superconducting metallic rhenium simultaneously as source and detector of \Re\ beta decay \cite{neutrino2010}.
After several years of attempts however metallic rhenium seems to be not fully compatible with the technical requirements of MARE and the focus of the community is shifting from this isotope to \Ho\ \cite{galeazzi}. 

\section{Electron Capture}
\begin{figure*}[!ht]
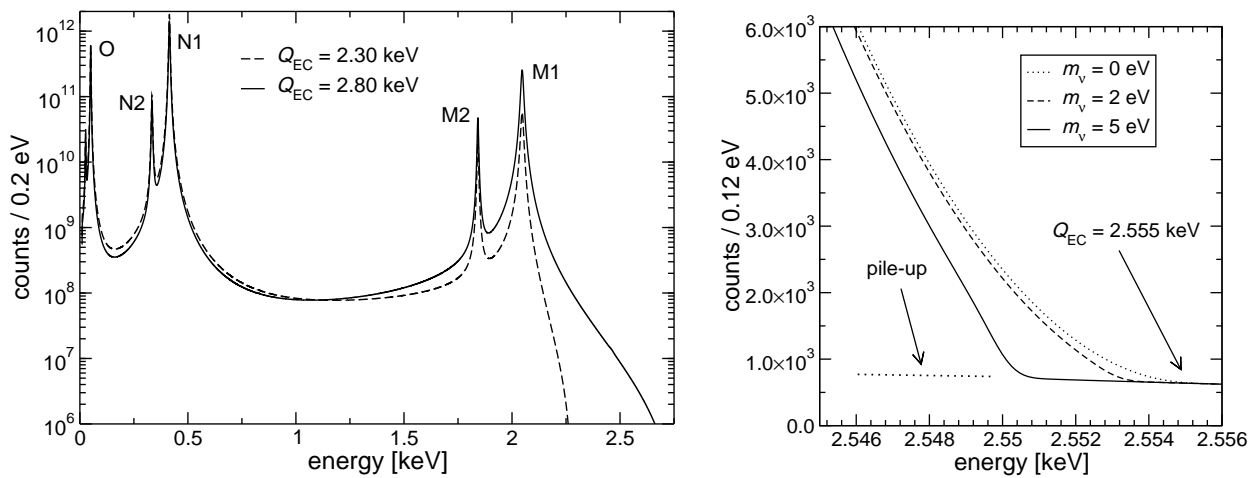

\begin{centering}
  \includegraphics[width=.505\textwidth]{olmio_cfr_Q-BW.eps}
\hspace{.02\textwidth}
  \includegraphics[width=.4\textwidth]{olmio_endpoint-BW.eps}
  \caption{\label{fig:spectrum}\Ho\ total absorption spectrum calculated for an energy resolution $\Delta E_{\mathrm{FWHM}}=1$\,eV and a pile-up fraction $f_{pp}=10^{-6}$ (left). Blow up of the spectrum close to the end-point (right).}
\end{centering}
\end{figure*}
Thirty years ago it was suggested that electron capture (EC) decays with 
low $Q$--values could be used as an alternative to single beta decay
for the direct determination of the electron neutrino mass. 

The most appealing proposal was made in 1982 by De Rujula and Lusignoli\,\cite{rujula83} and considers a calorimetric measurement of the \Ho\ spectrum, where all the de-excitation energy is recorded. \Ho\ decays by EC to $^{163}$Dy with a half life of about 4570 years and with the lowest known $Q$-value, which allows captures only from the M shell or higher. 
Because of the high specific activity of \Ho\ (about $2\times10^{11}$\,\Ho\ nuclei give one decay per second) the calorimetric measurements can be achieved by introducing relatively small amounts of \Ho\ nuclei in detectors whose design and physical characteristics are driven almost exclusively by the detector performance requirements and by the containment of the de-excitation radiation. 
So far, because of the very low source activities, in the few reported calorimetric measurements of \Ho\ decay the statistics was too low to allow the direct assessment of the $Q$-value, which was instead inferred from the ratios of the capture probabilities from different atomic shells. 
Presently the evaluations of the $Q$--value range from about 
2.3\,keV to 2.8\,keV,  
with a recommended value of $2.555\pm0.016$\,keV \cite{Rei10}.

In beta decays, the neutrino mass \mn\ can be measured because of the presence, 
in the decay energy spectrum, of the phase space factor for the antineutrino: 
$E_{\nu} \cdot p_{\nu} \simeq (Q-E_e) \cdot  \sqrt{(Q-E_e)^2-m_{\nu}^2}$, where $E_\nu$, $p_\nu$, and $E_e$ are the energy and momentum of the antineutrino, and the kinetic energy of the electron, respectively.

In the calorimetric measurement of an EC spectrum, the same neutrino phase space factor appears, with the total de-excitation energy $E_c$ replacing the kinetic energy of the electron $E_e$. 
The de-excitation energy $E_c$ is the energy released by all the atomic radiation emitted in the process of filling the vacancy left by the EC process, mostly electrons with energies up to about 2\,keV \cite{rujula83} (the fluorescence yield is less than $10^{-3}$). The calorimetric spectrum appears as a series of lines at the ionization energies $E_i$ of the captured electrons. These lines have a natural width of a few eV and therefore the actual spectrum is a continuum with marked peaks with Breit-Wigner shapes (Fig.\,\ref{fig:spectrum}). The distribution in de-excitation (calorimetric) energy $E_c$ is expected to be
\begin {eqnarray}
\label{E_c-distr}
{d \lambda_{EC}\over dE_c} &=& {G_{\beta}^2 \over {4 \pi^2}}(Q-E_c) \sqrt{(Q-E_c)^2-m_{\nu}^2} \;
\times  \\
&& \sum_i n_i  C_i \beta_i^2 B_i {\Gamma_i \over 2\,\pi}{1 \over (E_c-E_i)^2+\Gamma_i^2/4} \,, 
\nonumber
\end{eqnarray}
where $G_{\beta} = G_F \cos \theta_C$ (with the Fermi constant $G_F$ and the Cabibbo angle $\theta_C$), $E_i$ is the binding energy of the $i$-th atomic shell, $\Gamma_i$ is the natural width, $n_i$  is the fraction of occupancy, $C_i $ is the nuclear shape factor, 
$\beta_i$ is the Coulomb amplitude of the electron radial wave function (essentially, the modulus of the wave function at the origin) and $B_i$ 
is an atomic  correction for electron exchange and overlap. 
As for beta decay experiments, the neutrino mass sensitivity depends on the fraction of events close to the end-point, which in turn depends on the $Q$-value. Moreover, the closer the $Q$-value of the decay to one of the $E_i$, the larger the resonance enhancement of the rate near the end-point, where the neutrino mass effects are relevant. The resulting functional dependence of the end-point rate on the $Q$-value for the EC case is steeper than the $1/Q^3$ one observed for beta decay spectra.% 

The spectrum given in (\ref{E_c-distr}) is a first approximation. Recently, attention has been drawn to the possible modifications arising from the shake-up or shake-off processes, in which an outer second electron is excited to an unoccupied bound level or to a continuum level, respectively \cite{Robertson}. While the point is relevant, the magnitude of the effect will probably be much less than estimated in \cite{Robertson}. The reason is that the shake-up probabilities \cite{Carlson} used in \cite{Robertson} have been evaluated for a vacancy created in a process in which the nuclear charge is unchanged (such as photo-ionization or internal conversion). In these instances, the ``second electron'' that might be shaken up feels a potential increased by one positive unit of charge. In electron capture, contrariwise, the nuclear charge changes by minus one unit. In the potential felt by the outer second electron the decreased nuclear charge and that due to the absence of the captured electron add up to much less than unity. The shake-up probabilities in electron capture, therefore, are expected to be much smaller than in photo-ionization or similar processes. This well know fact \cite{Crasemann} has recently been rediscussed in \cite{ADR}.
An accurate calculation of these effects, as well as of the interference terms neglected in (\ref{E_c-distr}), is beyond the scope of this paper.

Despite the shortcomings of past experiments and the theoretical uncertainties, a calorimetric absorption 
measurement of \Ho\ EC decay currently seems the best approach for a new generation of neutrino mass experiments aiming at improving spectrometer sensitivity.
Nowadays, low temperature microcalorimeters reached the necessary maturity to  be used in a large scale experiment with good energy and 
time resolutions and are therefore the detectors of choice for a sub-eV holmium experiment.
It is also worth noting that in a calorimetric measurement the impact of the residual systematic theoretical uncertainties is limited, because the shape of the energy distribution near to its end-point is bound to be determined by the value of the neutrino mass.

\begin{figure}[!tb]
  \includegraphics*[width=.45\textwidth, clip=true,trim=0pt 0pt 0pt 0pt]{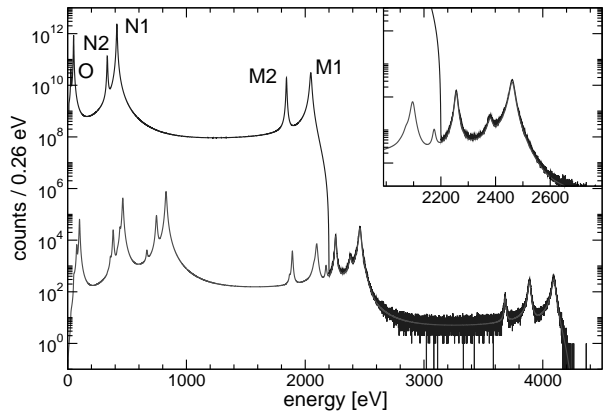}
  \caption{\label{fig:mc} Monte Carlo simulated full spectrum with $Q=2200$\,eV and for \Nev$=10^{14}$, \fpp$= 10^{-6}$, and \de\fwhm$=1$\,eV (dark line) and fitted pile-up spectrum (light line).}
\end{figure} 

\begin{figure}[!t]
\begin{center}
\resizebox{0.45\textwidth}{!}{\includegraphics*{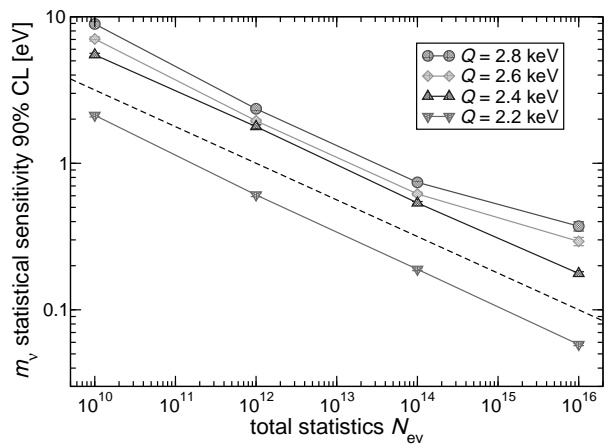}}
\end{center}
\caption{\label{fig:statistics}\Ho\ decay experiments statistical sensitivity dependence on the total statistics $N_{ev}$ for \de\fwhm$=1$\,eV, $f_{pp}=10^{-5}$, and no background.}
\end{figure}

In order to assess the statistical sensitivity of \Ho\ calorimetric experiments, a statistical analysis can be performed applying a frequentist Monte Carlo approach as outlined in \cite{nucciotti-epjc}. 
This numerical approach allows to include the pile-up, which is an important limiting factor for EC calorimetric measurements. 
The energy spectrum of pile-up events is given by the self-convolution of the calorimetric EC decay spectrum and extends up to $2Q$, producing therefore a background impairing the ability to identify the neutrino mass effect at the end-point (see Fig.\,\ref{fig:mc}).
The intensity of the pile-up spectrum is given by the two event pile-up probability $f_{pp}= \tau_R A_{EC}$, where $\tau_R$ is the time resolution and $A_{EC}$ is the EC activity in each detector.
This kind of statistical analysis shows that the total number of decays $N_{ev}$ is crucial for reaching a sub-eV neutrino mass sensitivity. In particular the sensitivity is proportional to $N_{ev}^{-1/4}$ as naively expected for a \mnsq\ sensitivity purely determined by statistical fluctuations. The current uncertainty on the $Q$-value translates to a factor 3 to 4 uncertainty on the neutrino mass sensitivity achievable by a  \Ho\ calorimetric  experiment. 
As an example, Fig.\,\ref{fig:statistics} shows that to reach a sub-eV statistical sensitivity on \mn\ using an experimental set-up  with an instrumental energy resolution \de\fwhm\ of about 1\,eV and a fraction of pile-up events $f_{pp}$ as low as $10^{-5}$ it is necessary to collect more than $10^{13}$ \Ho\ decays.

In addition to the HOLMES experiment, which is the subject of the present paper, there are currently two other experimental efforts aiming at measuring the neutrino mass using the EC decay of \Ho. The ECHo project plans to ion implant arrays of Metallic Magnetic Calorimeters with \Ho.  The ECHo collaboration has recently measured the \Ho\ decay spectrum with an energy resolution of about 8\,eV using two Metallic Magnetic Calorimeters which were ion implanted at ISOLDE (CERN) with a very small amount of \Ho\,\cite{Echo}. In the USA the NUMECS collaboration is critically assessing the various technologies required to produce, separate, and embed the \Ho\ isotope. NUMECS plans to measure the \Ho\ with arrays of transition edge calorimeters \cite{LosAlamos,Enge12}.

\section{The HOLMES experiment}
The HOLMES experiment is aimed at directly measuring the electron neutrino mass using the EC decay of \Ho\ and an array of low temperature microcalorimeters.
HOLMES was funded in 2013 by the European Research Council with an Advanced Grant awarded to Prof.\,S.\,Ragazzi. The Italian Istituto Nazionale di Fisica Nucleare (INFN) is the Host Institution for the grant which started on February 1$^{st}$ 2014 and will end on January 31$^{st}$ 2019. HOLMES continues the research program of the MARE project now focusing on the \Ho\ EC spectrum measurement.

HOLMES baseline experimental configuration has been defined through an extensive Monte Carlo statistical analysis based on the present knowledge of the \Ho\ decay parameters \cite{Lus11}:
in its optimal configuration HOLMES will collect about $3\times10^{13}$ decays with an instrumental energy resolution \de\ of about 1\,eV FWHM and a time resolution \taur\ of about 1\,\mus. For 3 years of measuring time \tm, this requires a total \Ho\ activity of about $3\times10^5$\,Bq. With an array of 1000 detectors, each pixel must contain an \Ho\ activity of about 300\,Bq which gives a \fpp\ of about $3\times10^{-4}$. The total activity is given by about $6.5\times10^{16}$ \Ho\ nuclei, or 18\,\mug, and each detector must therefore contain $6.5\times10^{13}$ \Ho\ nuclei. 
The choice of this configuration is mostly driven by the need for a statistics exceeding $10^{13}$ decays. In fact, for a fixed exposure $t_M \times N_{det}$, it pays out to increase the single pixel activity \Aec, and therefore \fpp, because the larger statistics \Aec\,$\times t_M \times N_{det}$ translates in a net statistical sensitivity improvement.

\begin{figure}[bt]
  \includegraphics[width=.45\textwidth, clip=true,trim=0pt 0pt 0pt 0pt]{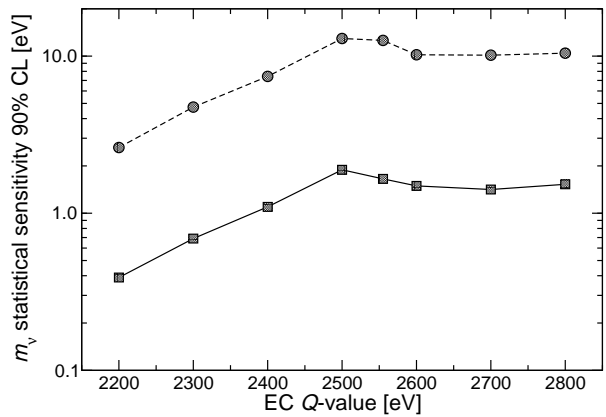}
  \caption{\label{fig:HOLMESsens}Monte Carlo estimate of HOLMES neutrino mass statistical sensitivity for \Nev$= 3\times10^{13}$ (lower curve) or $10^{10}$ (upper curve) and with \fpp$=3\times10^{-4}$, \de\fwhm$= 1$\,eV, and no background.}
\end{figure}

The 90\% C.L. statistical neutrino mass sensitivity estimated by the Monte Carlo simulation for the baseline HOLMES configuration and for various $Q$-values between 2.2 and 2.8\,keV spans from 0.4\,eV to about 1.8\,eV (lower curve in Fig.\,\ref{fig:HOLMESsens}). 
This mass sensitivity will be an important step forward in the direct neutrino mass measurement with a calorimetric approach as an alternative to spectrometry. It will also establish the potential of this approach to extend the sensitivity down to 0.1\,eV. A more detailed discussion of the achievable sensitivity is given in Section\,\ref{sec:sens}.

Thanks to the relatively short lifetime of \Ho\ -- as compared to \Re\ -- , the limited number of nuclei needed for a neutrino mass experiment can be introduced in the energy absorber of a standard low temperature detector. Therefore, HOLMES can leverage the microcalorimeters development for high energy resolution X-ray spectroscopy, with no need for a dedicated development of detectors like in the case of MARE with the metallic rhenium absorbers. In particular, the Transition Edge Sensors (TES) based microcalorimeters used for high resolution soft X-ray spectroscopy can be readily adapted to HOLMES specifications. Their technology is mature for a demanding neutrino physics experiment and presents many added values: 1) fast and full calorimetric energy measurements can be achieved in metallic absorbers, 2) small footprint kilo-pixel arrays are fabricated with well established photolithographic technologies, and 3) powerful SQUID multiplexing schemes for large detector arrays are available.

The baseline program of the HOLMES experiment is therefore to deploy an array of about 1000 TES based microcalorimeters each with about 300\,Bq of \Ho\ fully embedded in the absorber, and with the goal of energy and time resolutions as close as possible to 1\,eV and 1\,\mus, respectively.

HOLMES is characterized by 5 key tasks: 1) \Ho\ isotope production, 2) \Ho\ source embedding, 3) Single detector optimization and array engineering, 4) multiplexed read-out, and 5) data handling. 
In the following we describe these tasks and their current status.

\subsection{\Ho\ isotope production}
\label{sec:isotope}
$^{163}$Ho was discovered in 1960 in a sample of $^{162}$Er that was 
neutron activated in a nuclear reactor \cite{Naumann60}. Since its discovery, $^{163}$Ho has been 
produced in laboratory only for the purpose of studying nuclear and atomic properties. 
Since $^{163}$Ho  is not available off the shelf, a dedicated process must be set 
up to produce the amount needed by a neutrino mass experiment. 
The most challenging issue  is the achievement of the very high level of radio-purity 
required.

The $6.5\times10^{16}$ \Ho\ nuclei needed to carry out the experiment will be produced by neutron irradiation of \ero\ enriched in \Er\ through the reactions
\begin{equation}
^{162}\mathrm{Er}(\mathrm{n},\gamma)^{163}\mathrm{Er}\,\rightarrow\,^{163}\mathrm{Ho}\,+\,{\nu}_e  
\end{equation}
where the decay is an EC with a half-life of about 75\,min and the $\sigma(\mathrm{n},\gamma)$ is about 20\,barns for thermal neutrons.
The needed amount of enriched \ero\ depends on many parameters such as the neutron flux, the \Er\ and the -- yet unknown -- \Ho\ neutron capture cross sections (which govern the \Ho\ production and burn-up, respectively) \cite{Enge12,endf}, and the detector embedding efficiency (see next section).

The enriched \ero\ samples will be irradiated at the high-flux nuclear reactor of the Institut Laue-Langevin (ILL, Grenoble, France)  with a thermal neutron flux of about $1.3\times10^{15}$\,n/s/cm$^2$ \cite{ulli}. 
At this neutron high neutron flux reactor, neglecting the \Ho\ burn-up through the reaction  \Ho$(n,\gamma)^{164}$Ho, the \Ho\ production rate is approximately 50\,kBq(\Ho)/week/mg(\ero), for \ero\ enriched at 30 \% in \Er. 

The best neutron irradiation plan depends on the \Ho\ burn-up cross-section which is currently unknown. Assuming a burn-up cross-section of $\sigma(n,\gamma)\approx100$\,barns and a reasonable embedding efficiency of about 0.2\% (see \S\ref{sec:embedding}), for the purpose of the HOLMES research program we estimate a need for about 500\,mg of \ero\ enriched at 30\% in \Er\ to be irradiated at ILL for about 6\,weeks. 

Because of the presence of impurities  such as $^{164}$Er and $^{165}$Ho in the enriched \ero, along with \Ho, neutron irradiation will produce also $^{166m}$Ho, a $\beta$ decaying isomer with a half life of about 1200\,years which is potentially disturbing because of the background it causes below 5\,keV. 
The neutron activation produces also long living isotopes because of the neutron capture by other Er isotopes present in the enriched \ero\ samples. In particular, the irradiated sample could contain  $^{170}$Tm and  $^{171}$Tm with extremely high activities (the production rate at ILL is about few MBq/week/mg(enriched \ero))
which would prevent a safe handling.
The chemical separation of holmium from of the irradiated enriched \ero\ powder is therefore mandatory and will be performed at the Paul Scherrer Institute (PSI, Villigen, Switzerland) by means of ion exchange chromatography. 
The opportunity of performing a chemical purification step of the \ero\ sample also before neutron activation is still under evaluation.

In the past years, in the framework of the MARE project, many enriched \ero\ samples have been irradiated with the purpose of verifying the \Er\ $(n,\gamma)$ cross-sections, estimating the unknown \Ho\ $(n,\gamma)$ cross-section, and determining the best purification practices.
Some samples were irradiated at the Portuguese Research Reactor (RPI) at the Nuclear and Technological Institute (ITN) in Lisbon, which has a thermal neutron flux of about 10$^{13}$\,n/s/cm$^2$.
In 2013 about 25\,mg were purified at PSI and then irradiated at the ILL reactor for 50\,days: the resulting \Ho\ activity is expected to be about 10\,MBq. Small amounts of both irradiated and non-irradiated enriched \ero\ samples are presently being analyzed by means of Inductive Coupled Plasma Mass Spectroscopy (ICP-MS) at LNGS and at PSI.
Another 150\,mg of enriched \ero\ have been processed at PSI and the chemically purified \ero\ powder is now under irradiation at ILL.

\subsection{\Ho\ isotope embedding}
\label{sec:embedding}
In order to perform a calorimetric measurement of the EC spectrum, about $6.5\times10^{13}$ \Ho\ nuclei must be embedded in the absorber of each microcalorimeter. To avoid chemical shifts of the end-point, only holmium in the metallic chemical form must be introduced. 

The embedding system has therefore the purpose of implanting elemental \Ho\ in the detector absorbers discarding all other nuclides accidentally produced in the irradiation process which could be source of radioactive background or could cause an excess heat capacity in the detectors. This system is indeed a key component of the HOLMES experiment which is critical for achieving an high statistical sensitivity while minimizing the sources of systematic uncertainties. 
It consists of an ion implanter and a holmium evaporation chamber to produce the metallic target for the ion implanter source, and it will be set-up at the Genova INFN laboratory.

The ion implanter consists of four main components. A Penning sputter ion source \cite{ionsource}, a magnetic/electrostatic mass analyzer, an acceleration section, and an electrostatic scanning stage. The full system is being designed to achieve an optimal mass separation for \Ho. This will allow to separate \Ho\ from other trace contaminants not removed by chemical methods at PSI, such as $^{166m}$Ho. 
The implanter will be integrated in a vacuum chamber which will also allow a simultaneous gold evaporation to control the \Ho\ concentration and to deposit a final gold layer to prevent the \Ho\ from oxidizing.

The metallic cathode for the ion source will be made by a gold matrix containing metallic \Ho\ which will be produced in the holmium evaporation chamber.
In this latter chamber a tantalum heated Knudsen cell will be used for the reduction and distillation of metallic holmium from the
\hoo\ obtained by chemical separation from the irradiated \ero\ at PSI.
The \hoo\ powder (possibly containing traces of other lanthanides) will be thermally reduced at about 2000\,K using the reaction Ho$_2$O$_3$+2Y({\it met})$\rightarrow$2Ho({\it met})+ Y$_2$O$_3$.
At about 2000\,K holmium has by many orders of magnitude the highest vapor pressure and therefore will evaporate and deposit onto the substrate that will be used as cathode for the ion source. This process has been already tested  successfully in the past years in the framework of the MARE project \cite{elena-ltd}. Based on this experience a new thermally optimized cell has been designed to be installed in the vacuum chamber.
Also this vacuum chamber will allow a simultaneous gold evaporation to obtain a gold matrix with diluted \Ho.

After a testing and optimization period using $^{166m}$Ho -- traceable by means of $\gamma$ spectroscopy --, the system will be ready for implanting \Ho\ in the first detectors early in 2015.
The testing phase will allow to assess the efficiency of the entire process, from the neutron irradiation of the \ero\ powder to the detector absorber embedding. In fact, the embedding process introduces unavoidable losses of \Ho\ which may be particularly sizable in the ion source and in the ion beam scanning. We estimate the ion source efficiency to be about 5\%, while the ion beam scanning efficiency, which depends on both the array design and the ion beam cross-section, can be about 6\%. 
Allowing also for a combined efficiency of about 70\% for the other steps -- such as the holmium chemical separation and the metallic cathode production -- we expect a total isotope processing efficiency of about 0.2\%.
\subsection{TES detectors and array}
\begin{figure}[t]
\begin{centering}
 \includegraphics[width=0.45\textwidth ]{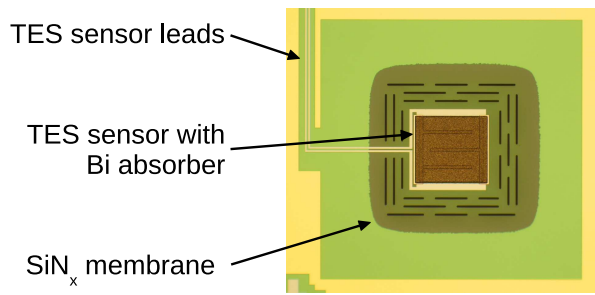}
 \caption{\label{img:tes}One TES with bismuth absorber fabricated by NIST.}
\end{centering}
\end{figure}
\begin{figure*}[!tb]
\begin{centering}
 \includegraphics[width=0.9\textwidth ]{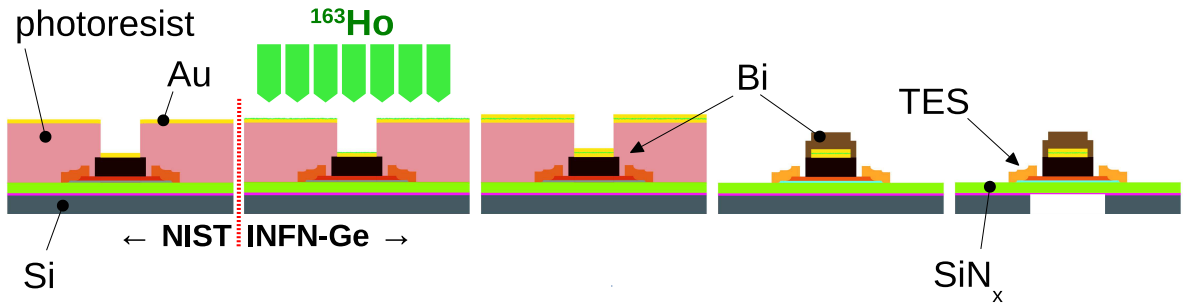}
 \caption{\label{img:fab}The two step TES fabrication process.}
\end{centering}
\end{figure*}
The detectors used for the HOLMES experiment will be Mo/Cu TES on \sinx\ membrane with bismuth absorbers (Fig.\,\ref{img:tes}). The TES microcalorimeters will be fabricated in a two step process. 
The first steps will be carried out at the National Institute for Standard and Technology (NIST, Boulder, Co, USA) \cite{nist-tes} where 
the devices will be fabricated up to the deposition of the bottom half of the absorber, i.e a 1.5\,\mum\ bismuth layer (Fig.\,\ref{img:fab}). 
The devices will be further processed in the Genova INFN laboratory (Fig.\,\ref{img:fab}). Here, the first step will be the deposition by means of the ion implanter of a thin (few 100\AA{}) layer of Au:\Ho, then the bismuth absorber will be completed with a deposition of a second 1.5\,\mum\ bismuth layer to fully encapsulate the \Ho\ source. GEANT4 simulations show that this bismuth 
 thickness is enough for fully containing  99.99997\% of the highest energy electrons ($\approx 2$\,keV) emitted in the \Ho\ decay.
The second step will be a Deep Reactive Ion Etching (DRIE) of the back of the silicon wafer to release the membranes with the TES microcalorimeters.

The relatively high concentration of holmium ($J=7/2$) could indeed cause an excess heat capacity due to hyperfine level splitting in the metallic absorber \cite{EnH05}. Low temperature measurements have been already carried out in the framework of the MARE project to assess the gold absorber heat capacity ($<150$\,mK), both with holmium and erbium implanted ions. Those tests did not show any excess heat capacity, but further more sensitive investigations will be carried out \cite{HoAu}. If necessary, dilution of the implanted \Ho\ concentration will be achieved by co-evaporation of gold during the implantation (see \S\ref{sec:embedding}).

The TES array is presently being designed with the aim of 
matching as close as possible the baseline goal of an
energy resolution \de\fwhm\ of about 1\,eV at the spectrum end-point and a time resolution $\tau_R$ of about 1\,\mus. 
This requires an optimal thermal design of all detector components. To minimize the stray electrical inductance $L$ which limits the pulse rise time, the TES will be arranged in $2\times32$ sub-arrays. This arrangement allows also to maximize the geometrical filling factor and therefore the \Ho\ implantation efficiency. 

As discussed in the next section, together with the detector itself also its read-out plays a crucial role for achieving the specified performances.

\subsection{Read-out and data acquisition}
\label{sec:mux}
The signals from the 1000 microcalorimeters will be read-out with a multiplexed system which is critical to preserve the performances of the individual detectors, especially in terms of available signal bandwidth, i.e. time resolution $\tau_R$. The baseline for the HOLMES experiment, as outlined in the proposal, is to use a Code Division Multiplexing (CDM, a special version of Time Domain Multiplexing) which has been developed by NIST \cite{nist-cdm}.

\begin{figure}[b]
  \includegraphics[width=.45\textwidth]{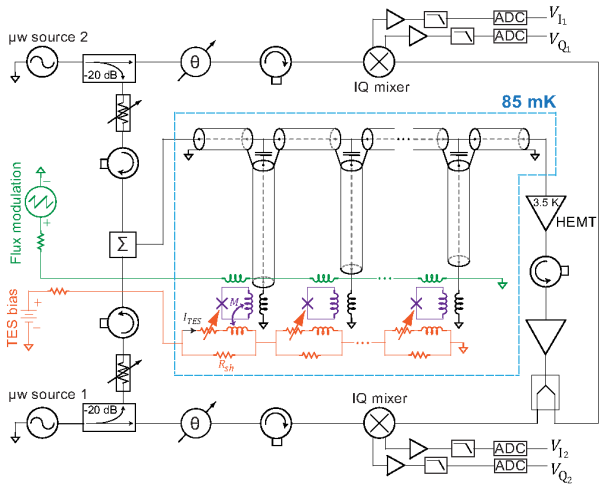}
  \caption{\label{fig:mumux}Circuit schematic for a two channel microwave multiplexed read out of a TES array (from \cite{nist-mumux}).}
\end{figure}

Recent advancements on microwave multiplexing ($\mu$MUX) suggest that this is the most suitable system for HOLMES, since it gives a larger bandwidth for the same multiplexing factor (number of multiplexed detector signals).
It is based on the use of rf-SQUIDs as input devices with flux ramp modulation (Fig.\,\ref{fig:mumux}) \cite{nist-mumux}. The modulated rf-SQUID signals are read-out coupling the rf-SQUID to superconducting LC resonators in the GHz range and using the homodyne detection technique. Tuning the LC resonators at different frequencies it is straightforward to multiplex many RF carriers.

Although there is no practical demonstration of the feasibility of this approach at the level required by HOLMES, the TES microcalorimeter $\mu$MUX principle has been demonstrated in \cite{nist-mumux} with two channels. We expect to exploit the TES $\mu$MUX scalability by
leveraging many of the technological advancements of similar instruments such as the ARCONS large multiplexed array of Kinetic Inductance Detectors for single photon imaging\,\cite{arcons} or the multiplexed arrays of TES bolometers with flux ramp modulation of MUSTANG2\,\cite{Dicker}. 

The $\mu$MUX is suitable for a fully digital approach based on the Software Defined Radio (SDR) technique. The comb of frequency carriers is generated by digital synthesis in the MHz range and up-converted to the GHz range by $IQ$-mixing. The GHz range comb is sent to the cold $\mu$MUX chips coupled to the TES array through one semirigid cryogenic coax cable, amplified by a cryogenic low noise High Electron Mobility Transistor (HEMT) amplifier, and sent back to room temperature through another coax cable.
The output signal is down-converted by $IQ$-mixing, sampled with a fast A/D converter, and digital mixing techniques are used to recover the signals of each TES in the array ({\it channelization}). 

One of the best available systems for realizing an SDR is the Reconfigurable Open Architecture Computing Hardware (ROACH2) board equipped with a FPGA Xilinx Virtex6 \cite{roach2}. ROACH2 has been developed in the framework of CASPER (Collaboration for Astronomy Signal Processing and Electronic Research) and is a fully open system which has the support of a large community of developers. 
The complete acquisition system is made by the ROACH2 board coupled to DAC (for the comb generation), ADC (for signal digitization) and IF (for signal up- and down-conversion) boards.

The FPGA installed on the ROACH2 board will be programmed to perform {\it channelization} and real-time digital signal processing. 
For this purpose a new software is being developed to test and optimize the algorithms required for optimal pulse height estimation and rise-time pile-up detection.

The number of TES signals which can be multiplexed on one ROACH2 board -- the multiplexing factor -- is set by the bandwidths of both the A/D converter and the TES signal. It can be shown that the multiplexing factor is approximately given by $0.01 f_\mathrm{ADC} \tau_{rise}$, where $f_\mathrm{ADC}$ is the ADC sampling frequency and  $\tau_{rise}$ is the TES signal rise time (10-90\%). With the ADC boards currently available for ROACH2 (550\,MS/s, 12\,bit), an acceptable trade off is a multiplexing factor of about 50 and TES signals with a
rise time $\tau_{rise}$ of about 10\,\mus. Increasing the multiplexing factor or decreasing the rise time requires faster A/D converters.
To assess the time resolution $\tau_R$ which can be achieved with this configuration, we have developed a code to test various algorithms for pile-up identification on simulated realistic TES signals and noise. These simulations suggest that a time resolution $\tau_R$ of at least 2\,\mus\ should be achievable for signals with a rise time $\tau_{rise}$ of about 10\,\mus\ (10-90\%) and a signal-to-noise ratio corresponding to
an energy resolution at the \Ho\ $Q$-value of about \de\fwhm\ of 2\,eV after optimal signal filtering.
These resolutions closely match the baseline specifications of the HOLMES experiment and there is even some margin for improvements.

Work is in progress to optimize the microwave SQUID multiplexed read-out and experimentally assess its sensitivity: preliminary results on both the SQUID noise and the cross-talk are very promising  \cite{bennett}.

\subsection{Signal processing and data handling}
The high data throughput caused by about $3\times10^5$ \Ho\ decays per second prevents from saving all fully digitized events and 
calls for an unconventional signal processing strategy to minimize the size of the data to be analyzed and stored off-line. 
This will be obtained  
by exploiting the FPGAs on the ROACH2 boards to perform most of the signal processing in real-time. 
Fully digitized event records will be saved to disk for off-line analysis only during the commissioning phase and on a regular basis to check the stability of the detector response and the effectiveness of the filtering algorithms. 
During the commissioning phase the parameters for the real-time signal processing will be determined on a channel-by-channel basis. Commissioning data will be recorded with the lowest energy threshold and will amount to about $5\times10^{10}$ events which will be used to fully characterize the \Ho\ decay spectrum, to tune the pile-up resolving algorithms, to identify critical event categories where environmental noise has not been properly filtered, and to determine the optimal filter. 
During regular data taking an energy threshold will be applied and for each event only the relevant calculated parameters will be saved for off-line analysis.
In this way we estimate that the data accumulated by HOLMES will be limited to about 150\,TByte for 3\,years of data taking.

The reduced data sample will be used to produce the final energy spectrum needed for the end-point analysis. It will result from energy calibration, gain stability correction, pile-up and spurious pulse rejection on a channel-by-channel basis. Background events caused by cosmic rays and environmental noise or radioactivity will be reduced by applying an anti-coincidence cut among detectors. Individual energy calibrated spectra will be then summed up to give the final total spectrum. The end-point statistical analysis together with the systematic uncertainty investigation will be performed with the usual Bayesian and/or frequentist software tools.

It is worth noting that, during data taking, HOLMES will not need an external radiation source for energy calibration and gain instability correction. The M and N capture peaks provide a constant energy reference which will be used for these purposes. In this respects, HOLMES detectors are self-calibrating and it will be therefore straightforward to obtain the detector energy response function needed for the end-point analysis. The high statistics of the measurement will also allow a very precise determination of all atomic parameters (natural widths, asymmetries, branching ratios and peak positions) that significantly help to reduce systematic uncertainties. 
\section{Sensitivity}
\label{sec:sens}
One of the goals of HOLMES is to identify and characterize the systematic uncertainties which might
deteriorate the final neutrino mass sensitivity.
Nevertheless also the \mn\ statistical sensitivity  shown in Fig.\,\ref{fig:HOLMESsens} may be 
threatened by unmatched detector performance. 

\begin{figure}
  \includegraphics[width=.45\textwidth, clip=true,trim=0pt 0pt 0pt 0pt]{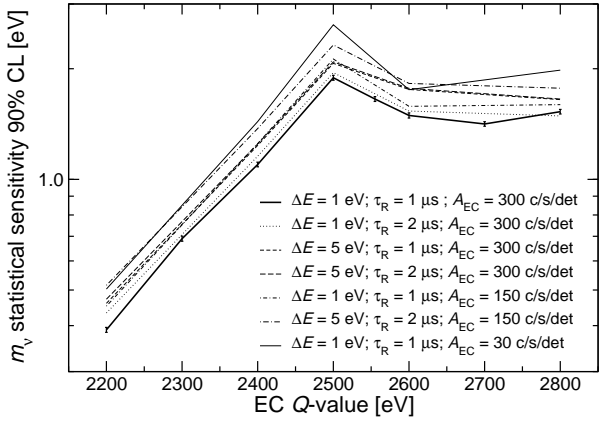}
  \caption{\label{fig:HOLMES-sens-bad} Monte Carlo estimates of the effect on HOLMES baseline statistical sensitivity of different deviations from the baseline configuration and planned performance.}
\end{figure}

Fig.\,\ref{fig:HOLMES-sens-bad} shows how the statistical sensitivity may worsen because of various combinations of poor energy and time
resolutions or reduced implanted \Ho\ activities, with the experimental exposure $t_M \times N_{det}$ kept constant at 3000\,year$\times$detector. This plot shows the tolerance of HOLMES for performances which are worse than the design ones.
In particular, the presently expected energy and time resolutions of about 2\,eV and 2\,\mus\ (see Section\,\ref{sec:mux}) would cause
a only negligible deterioration of the sensitivity.

\begin{figure}
  \includegraphics[width=.45\textwidth, clip=true,trim=0pt 0pt 0pt 0pt]{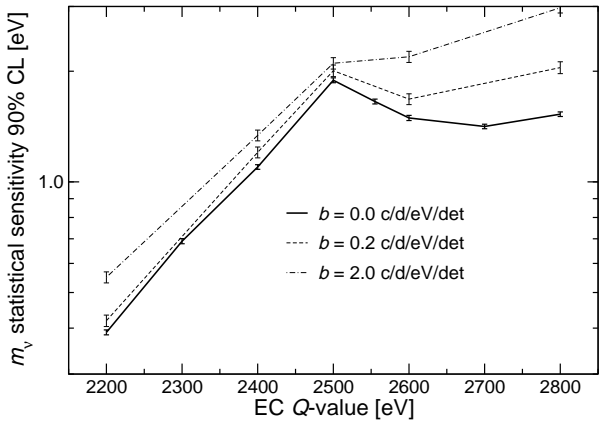}
  \caption{\label{fig:HOLMES-bkg} Monte Carlo estimates of the effect of various background levels on HOLMES baseline statistical sensitivity.}
\end{figure}

Radioactive background might be another factor limiting the statistical sensitivity.  
A constant background $b$ is indeed negligible as long as it is much lower than the pile-up spectrum, i.e. when $b\ll$\Aec\fpp$/2Q$ or about 1.5\,count/eV/day/detector for HOLMES baseline configuration. Fig.\,\ref{fig:HOLMES-bkg} shows to what extent a constant background between 0 and $2Q$ can reduce HOLMES baseline sensitivity: the two curves are for constant backgrounds $b$ comparable to 1.5\,count/eV/day/detector. 

Background in an experiment like HOLMES is caused by cosmic rays -- both muons crossing the microcalorimeter absorber and electromagnetic showers caused by nearby interactions of muons --, by environmental radiation -- such as environmental $\gamma$s or X-rays and Auger electrons from surrounding materials --, and by radioactive contaminations inside the microcalorimeter absorbers.   

Minimum ionizing muons crossing a 3\,\mum\ thick bismuth absorber deposit about 5\,keV on the average and the rate at sea level for a $200\times200$\,\mum$^2$ horizontal absorber is of about one muon per day: these numbers roughly give a background of about $10^{-4}$\,count/eV/day/detector which is confirmed also by Monte Carlo simulations. Actual HOLMES muon background should be even lower because of the concrete overburden at the HOLMES site.
 
The MIBETA experiment with \agre\ microcalorimeters \cite{mibeta} observed a flat background of about $1.4\times10^{-4}$\,count/eV/day/detector between 0 and about 5\,keV. 
Geant4 Monte Carlo simulations show that -- thanks to the combined effect of smaller volume and larger $Z$ of the bismuth absorbers --, the contribution to the counting rate below 10\,keV for incident $\gamma$s of any energy between 10\,keV and 3\,MeV is lower in HOLMES than in MIBETA. 
Therefore in HOLMES the level of the background due to environmental $\gamma$s should be lower than the total background rate measured by MIBETA. 
Contributions to the background due to X-rays and Auger electrons from surrounding materials should be comparable for the two experiments.
All the above suggests that HOLMES should have a negligible background much smaller than 1.5\,count/eV/day/detector, but a precise assessment will be made running an array of TES microcalorimeters without \Ho\ in the bismuth absorbers. If necessary, the cosmic ray background can be reduced by mounting the array vertically or by using a small veto detector just above the array. Moreover, past measurements done in the framework of MARE showed that a 40\,mm shield made out of radio-pure copper positioned around the microcalorimeter array can give a sensible reduction of the environmental background even at shallow depth.

The relatively low sensitivity of HOLMES to cosmic rays and environmental radioactivity is a consequence of the high pile-up rate which
is expected in the baseline experimental configuration. Of course, a lower detector activity \Aec\ would make 
the background issue more relevant by reducing the pile-up probability \fpp.
In the unlikely event that the external background turns out to limit the achievable \mn\ statistical sensitivity, HOLMES can be moved to the INFN Gran Sasso Underground Laboratory (LNGS).

HOLMES sensitivity is more likely challenged by background counts caused by radioactive contaminations internal to the microcalorimeters.
As discussed previously, radioactive isotopes produced along with \Ho\ might be incorporated in the microcalorimeter absorber during the embedding procedure, although the chemical separation at PSI and the mass selection in the \Ho\ implanter system should minimize this risk.
The most worrisome isotope is $^{166m}$Ho: Geant4 Monte Carlo simulations show that its contribution to the background level in the region of interest is about $10^{-11}$\,count/eV/day/($^{166m}$Ho nucleus). The production rate of $^{166m}$Ho depends on many factors, such as for example the amount of $^{164}$Er and $^{165}$Ho, in the starting \ero\ powder or the importance of processes involving high energy neutrons. 
Under the pessimistic assumption to produce a $^{166m}$Ho contamination at the per mill level, we would expect a background of about 1\,count/eV/day/detector. This sets a strong constraint on the tails of the resolution function of the mass separation stage in the \Ho\ implanter system and shows how critical the embedding system is for the success of HOLMES.

\section{Conclusions}
Presently all the systems required for the HOLMES experiment are being designed and set-up. A cryogenic test-bed instrumented for TES arrays microwave readout is being set up at University of Milano-Bicocca laboratory. Measurements on the first prototype detectors with implanted \Ho\ are expected for mid 2015. These preliminary measurements will yield important informations such as the $Q$-value and the isotope embedding efficiency.
A one month preliminary measurement with only 16 detectors identical to the final ones is planned for late 2016. Such a measurement will provide the sensitivity shown by the upper curve in Fig.\,\ref{fig:HOLMESsens}.
With the measurements performed during its preparation phase, HOLMES will measure with great precision all the atomic and nuclear parameters of the \Ho\  EC decay. The final measurement with the full array is expected to start in late 2017.
%%%%%%%%%%%%%%%%%%%%%%%%%%%%%%%%%%%%%%%%%%%

%%%%%%%%%%%%%%%%%%%%%%%%%%%%%%%%%%%%%%%%%%%%%%%%
%% BACKMATTER
%%%%%%%%%%%%%%%%%%%%%%%%%%%%%%%%%%%%%%%%%%%%%%%%

\section{Acknowledgments}
The HOLMES experiment is funded by the European Research Council under the European Union's Seventh Framework Programme (FP7/2007-2013)/ERC Grant Agreement no. 340321.
We also acknowledge support from INFN for the MARE project, from the NIST Innovations in Measurement Science program for the TES detector development, and from Funda\c{c}\~{a}o para a Ci\^{e}ncia e a Tecnologia (PTDC/FIS/116719/2010) for providing the enriched \ero\ used in preliminary \Ho\ production by means of neutron irradiation.

\end{document}